\documentclass[%
 reprint,
 amsmath,amssymb,
 aps,
]{revtex4-2}

\usepackage{graphicx}% Include figure files
\usepackage{dcolumn}% Align table columns on decimal point
\usepackage{bm}% bold math
\usepackage{color}
\usepackage{float}

\begin{document}

\preprint{APS/123-QED}

\title{Unraveling higher-order dynamics in collaboration networks}% Force line breaks with \\ %Temporal Evolution of the Structure of Collaboration Hypergraph
 
%\affiliation{Instituto de Física Interdisciplinar y Sistemas Complejos IFISC (CSIC-UIB), Palma de Mallorca, Spain.\\
%  Department of Mathematical Sciences ``Giuseppe Luigi Lagrange'' (DISMA), Politecnico di Torino, C.so Duca degli Abruzzi 24, 10129, Torino, Italy.\\
 % Ciessehacca\\
 % ISI Foundation, Turin, Italy.\\
 % Department of Physical Sciences, Indian Institute of Science Education and Research, Kolkata, India.\\
 % University of Urbino, Piazza della Repubblica 13, Urbino, Italy.
  %}

\author{David Abella$^{1}$}
\author{Piero Birello$^{2}$}
\author{Leonardo Di Gaetano$^{3}$}
\author{Sara Ghivarello$^{4}$}
\author{Narayan G. Sabhahit$^{5}$}
\author{Christel Sirocchi$^{6}$}
\author{Juan Fernández-Gracia$^{1}$}

\affiliation{$^{1}$Instituto de Física Interdisciplinar y Sistemas Complejos IFISC (CSIC-UIB), Palma de Mallorca, Spain.}
\affiliation{$^{2}$Department of Mathematical Sciences Giuseppe Luigi Lagrange (DISMA), Politecnico di Torino, 10129 Turin, Italy.}
\affiliation{$^{3}$Department of Network and Data
Science, Central European University, 1100 Vienna, Austria. }
\affiliation{$^{4}$Institute for Scientific Interchange (ISI), 10126 Turin, Italy.}
\affiliation{$^{5}$Department of Physical Sciences, Indian Institute of Science Education and Research, Kolkata, India.}
\affiliation{$^{6}$Department of Pure and Applied Sciences, University of Urbino, Urbino, Italy.}
%  \email{Second.Author@institution.edu}
% \affiliation{%
%  Authors' institution and/or address\\
%  This line break forced with \textbackslash\textbackslash
% }%

\begin{abstract}
The interactions between individuals play a pivotal role in shaping the structure and dynamics of social systems. Complex network models have proven invaluable in uncovering the underlying mechanisms that govern the formation and evolution of these systems. However, conventional network representations primarily emphasize pairwise interactions, represented as edges in the network. In reality, many social interactions occur within groups rather than individual pairs. To capture this crucial aspect, higher-order network representations come into play, especially to describe those complex systems that are inherently composed of agents interacting with group dynamics. Despite recent research advancements in exploring temporal higher-order networks in various systems, our understanding of collaboration networks remains limited. Specifically, there is a lack of knowledge regarding the patterns of group interactions within scientific collaborations. How do groups form and evolve in this context? In this study, we aim to delve into the temporal properties of groups within collaboration networks. Our investigation focuses on uncovering the mechanisms that govern the global, group, and individual-level dynamics, shedding light on how individuals collaborate and how groups form and disband over time. By studying these temporal patterns, we take a significant stride forward in comprehending the intricate dynamics of higher-order interactions within human collaboration systems.

\end{abstract}

%\keywords{Suggested keywords}%Use showkeys class option if keyword
                              %display desired
\maketitle

%\tableofcontents

\section*{Introduction}

% motivation for why we want to study the collaboration pattern and the importance of studying social systems... (cite science of science papers and books?)\\
% \\
% Networks are pretty good at that..\\
Scientific progress is intricately linked to collaborative efforts, where researchers from diverse backgrounds come together to generate new knowledge and push the boundaries of scientific understanding~\cite{uzzi2013atypical,olechnicka2019geography}. Understanding the collaboration patterns within scientific communities is essential for unraveling the underlying structures and dynamics of social systems~\cite{bozeman2004scientists,glanzel2005analysing,abbasi2011identifying,adams2013fourth,shi2015weaving}. The field of science of science has emerged to investigate these phenomena, aiming to uncover the principles and mechanisms that govern scientific collaborations and their impact on knowledge production~\cite{fortunato2018science}. The motivation for studying collaboration patterns lies in the numerous benefits it offers. First, understanding how researchers form and maintain collaborations can inform policy decisions and resource allocation, facilitating the creation of effective support structures for scientific communities. Second, studying collaboration patterns can shed light on the emergence of scientific breakthroughs and the diffusion of knowledge, enabling the identification of influential individuals and key drivers of scientific progress. Finally, analyzing collaboration networks contributes to the broader understanding of social systems, highlighting the similarities and differences between scientific collaborations and other domains of human interaction~\cite{zeng2017science,wang2021science,wuchty2007increasing}.

In the past few decades, network science has emerged as a very important and useful tool for understanding the large-scale organization of complex physical, biological, and social systems~\cite{newman2018networks,albert2002statistical,newman2003structure,latora2017complex, chowdhary2023temporal}. A social system is usually modeled as a complex network called a social network, with nodes representing people or organizations and edges representing their relations and interactions with others. %. 

Since the early beginning of the century, and shortly after the seminal papers of Watts \& Strogatz~\cite{Watts1998} and Barabási~\cite{doi:10.1126/science.286.5439.509} that revived network science, scientists have been studying collaboration networks discovering that these networks showed the small-world property, high clustering~\cite{PhysRevE.64.016131,newman2001scientific,newman2001structure,barabasi2002evolution,guimera2005team} and a rich community structure~\cite{girvan2002community}. These studies conceptualized author collaboration networks by using simple weighted networks where authors were linked to one another by edges endowed with a weight equal to the number of papers they had written together. In subsequent papers, bipartite structures were taken into account, where nodes of two types appear, representing authors and papers~\cite{PhysRevE.70.036106}. In those studies, the author's collaboration network emerges as a simple projection of the bipartite network~\cite{Projection} over the papers. In these kinds of networks, edges are present only between pairs of nodes of different kinds, representing an author contributing to a paper.

%However, this static representation of social networks loses an important aspect of mechanism, link activation, that greatly affects the dynamics of agents interacting via the network. Temporal Network framework \cite{holme2012temporal} has been successful in modeling dynamical social systems uncovering hidden dynamical patterns in the structure of interaction.Unlike traditional static social networks, which represent relationships at a single point in time, temporal social networks provide a more comprehensive view of social dynamics by considering the temporal dimension.
Unlike traditional static social networks, which represent relationships at a single point in time, temporal social networks provide a more comprehensive view of social dynamics by considering the temporal dimension~\cite{holme2012temporal} opening up pathways to quantify and understand how groups of heterogenous size coalesce or disengage to give rise to the large scale structure \cite{korbel2023homophily}. Such a generalization of network representation considers mechanisms like link activation, which greatly affect the dynamics of agents interacting via the network. Temporal network formalism has been successful in uncovering Bursty phenomena in Human dynamics~\cite{karsai2018bursty, unicomb2021dynamics, hiraoka2020modeling}.

%Scientific collaboration Networks are one such important class of social networks, where the nodes represent the authors and an edge between the nodes represents a paper that they have collaborated with. The network representation of the collaboration data has been successful in uncovering the large-scale dynamical structure of the collaboration pattern\cite{newman2001structure,newman2001scientific,barabasi2002evolution}, which helps us to predict the emergence of innovation and scientific output \cite{fortunato2018science,wang2021science} in our society. 

While these studies have been very helpful in providing new and novel insights into the collaboration pattern, they suffer from the drawback that author collaboration networks, by definition, are only able to incorporate pairwise interactions~\cite{benson2016higher}. Hence, to bridge this gap in understanding collaboration patterns within scientific communities, more recently, higher-order representations like hypergraphs have been employed to encode group interactions in complex systems. These higher-order representations have proven valuable in unraveling the mechanisms underlying various phenomena, allowing for a deeper comprehension of collaborative dynamics and knowledge exchange among researchers~\cite{battiston2020networks,bianconi2021higher,battiston2021physics,torres2021and, boccaletti2023structure}. In recent years, endeavors have been made to  model the co-authorship data as a simplicial complex which revealed a simplicial closure event, where a group of nodes evolved until they finally became a part of a  higher-order structure~\cite{Patania2017TheSO, benson2018simplicial}. Hypergraph models were used to represent scientific output~\cite{lung2018hypergraph, di2023percolation,petri2018simplicial} and to suggest the author collaboration data can be modeled as a multilayer hypergraph where each layer is a d-uniform hypergraph (papers with the same number of collaborators), providing novel microscopic insights about the diversity of researchers within different teams and their relationships with others~\cite{vasilyeva2021multilayer}.  Previous studies have also addressed the heterogeneous dynamics of group interactions, finding differences between work and informal settings~\cite{cencetti2021temporal}.

Many of the existing works investigate static collaboration hypergraphs however, the dynamical behavior of such collaborations remains largely unexplored. Our results complement these recent research directions and build on shedding more light on the dynamics of collaboration hypergraphs.
%In this paper, we aim to address this gap by examining the dynamics of collaboration hypergraphs, complementing recent research directions in the field. 
Our study begins by inspecting the global temporal properties of scientific collaborations, revealing a gradual and significant increase in collaboration size. We proceed with the study of co-authorship from the perspective of individual researchers, revealing that the average order of collaboration for a single author increases throughout their career, while the average number of new collaborations formed decreases over time. Additionally, we investigate the tendencies of groups of authors involved in an article within a given year to either gain or lose members in subsequent publications. Notably, our findings indicate that groups exhibit a higher probability of remaining the same in terms of membership, but over time, there is an increased likelihood of both losing and adding members, highlighting a growing dynamism over the years. Furthermore, we explore the relationship between group size and the tendency to lose members, revealing that groups of size 7 display a lower likelihood of changing their size, suggesting a point of stability within the aggregation and disaggregation dynamics. Throughout the work, we provide intuitive insights and interpretations of our results that can inform future work aiming to develop a model for understanding this phenomenon in greater detail. 
%In many real-world social systems, the interactions can be of "higher order" in nature. More recently, much attention has been given to bridging this gap \cite{} (some hoi review). Our goal in this manuscript is to shed more light on the temporal dynamics of real-world hypergraphs. To that end, we construct a co-authorship hypergraph and study the structural and dynamical properties of the system. More specifically, we answer the following questions in our manuscript:

\section*{Results}

We consider a temporal higher-order network dataset on research collaborations in the form of published articles. The dataset consists of a sequence of timestamped hyperlinks (collaborations) where each simplex is a set of nodes (authors). 
In particular, we study scientific production for over 65 years in the research domain of Geology. Further details can be found in the \textit{Data and Code Availability} section.
We perform some data pre-processing in order to limit our analysis from 1950 to 2015, considering only authors that started their careers in this time window.
Our aim is to investigate the temporal aspects of higher-order interactions in the scientific community under observation~\cite{cencetti2021temporal,badie2022directed,ceria2023temporal}. We first investigate the historical behavior of groups over the years, and how the higher-order topology of the dataset has evolved over time. Furthermore, by means of a higher-order perspective, we investigate the patterns of careers of researchers~\cite{vaccario2020mobility} and the aggregation and disaggregation dynamics of groups. 
In the following, we will refer to \textit{size} as the dimension of the set of nodes that form an hyperedge (i.e., a collaboration of two authors has size two). We will refer to \textit{order} of the interaction as the hyperedge size minus one, accordingly to the Network Science community convention. 

%This is a temporal higher-order network dataset, which here means a sequence of timestamped simplices where each simplex is a set of nodes. In this dataset, nodes are authors and a simplex is a publication marked with the "Geology" tag in the Microsoft Academic Graph. Timestamps are the year of publication. The projected graph is a weighted undirected graph representing how many times each pair of nodes co-appears in a simplex. We restricted to simplices that consist of at most 25 nodes.
\subsection*{Higher-order scientific collaborations over years}
\begin{figure}
    \centering
    \includegraphics[width=\linewidth]{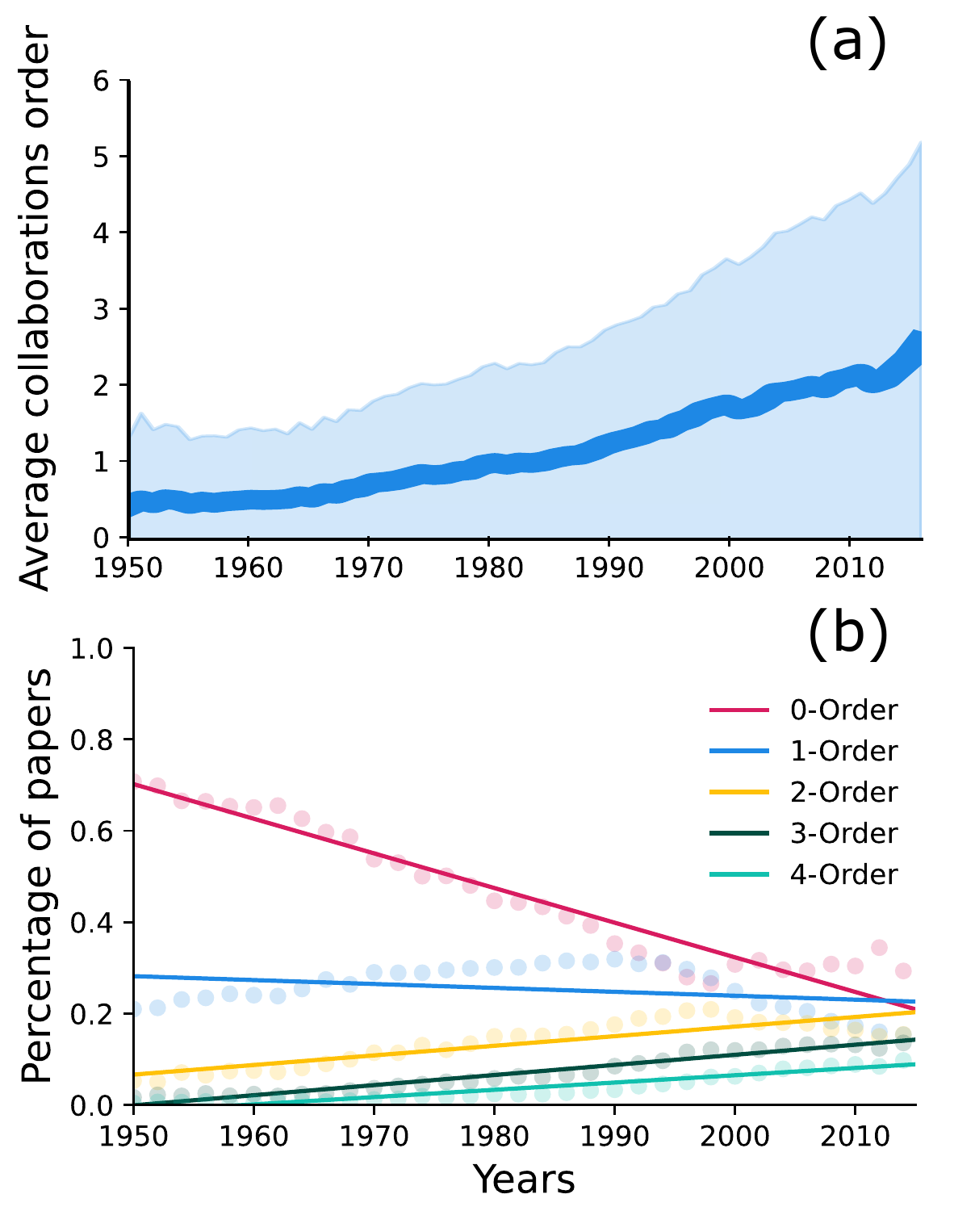}
    \caption{Orders of scientific collaborations over the years. (a) Average order (number of authors -1) of collaborations over the years (solid line) with a confidence band of one standard deviation. The average number of authors involved in scientific collaborations increases over time. (b) Fraction of papers of a given order over the years. Solid lines represent the linear fit of empirical observation (scatter points). Small scientific groups become less common over time while large groups become more common.}
    % for fig 1 a it would be nice to show median, 5 and 95 percentiles. For fig.1b would be nice to give the results of the fitings, at least in the SM
    \label{fig:panel_1}
\end{figure}

We first inspect the global temporal properties of scientific collaborations~\cite{bordons2000collaboration,newman2001structure,kumar2015co,aguirre2023methodology}. To this extent, we analyze how the average order of hyperlinks changes over the years, considering snapshots of hypergraphs made of works of a certain year and computing the average order of their collaborations, see Fig.~\ref{fig:panel_1}(a). We observe that the number of authors in collaborations has been increasing over the whole observed period. Specifically, the size of collaborations in 2015 has been approximately triplicated since 1950.
This trend can be caused by the technological development of the last decades that has made large collaborations more feasible, thanks to more accessible communications, improved transportation, and more connected scientific networks worldwide. 

A more detailed view of collaborations over time is portrayed in Figure~\ref{fig:panel_1}(b). Here we show the fraction of hyperlinks of a certain order over the years. The coauthorship network manifests a decreasing trend for 0-order (single author) and 1-order (2 authors) works. On the contrary, large research groups have become more relatively frequent over time, compatible with previous observations. 
The phenomenon of collaboration's size growth over the years arises the need for an understanding of group enlargement and aggregation dynamics, and of how a single researcher forms their connection in science throughout their career. We present our results in the \textit{Higher-order description of scientific career} and the \textit{Temporal dynamics of collaborations} sections.

\subsection*{Higher-order description of scientific career}
The frequency of the order of collaboration is an aggregated observable that does not give us information about the author-wise characteristics. Also, its evolution has been analyzed on a macroscopic time scale that differs from the characteristic lifespan of a researcher's activity. To fill this gap, we now investigate the local properties of nodes over their activity period, to seek patterns during the career of researchers. We detect the beginning and the end of each author's career. We keep track of all collaborations the author is involved in during the detected lifespan of the node. We are interested in finding recurrent patterns in authors' careers, in terms of engagement in multi-authors collaborations, preferential size of collaborations, and network neighborhood formation. 

\begin{figure}
    \centering
    \includegraphics[width=\linewidth]{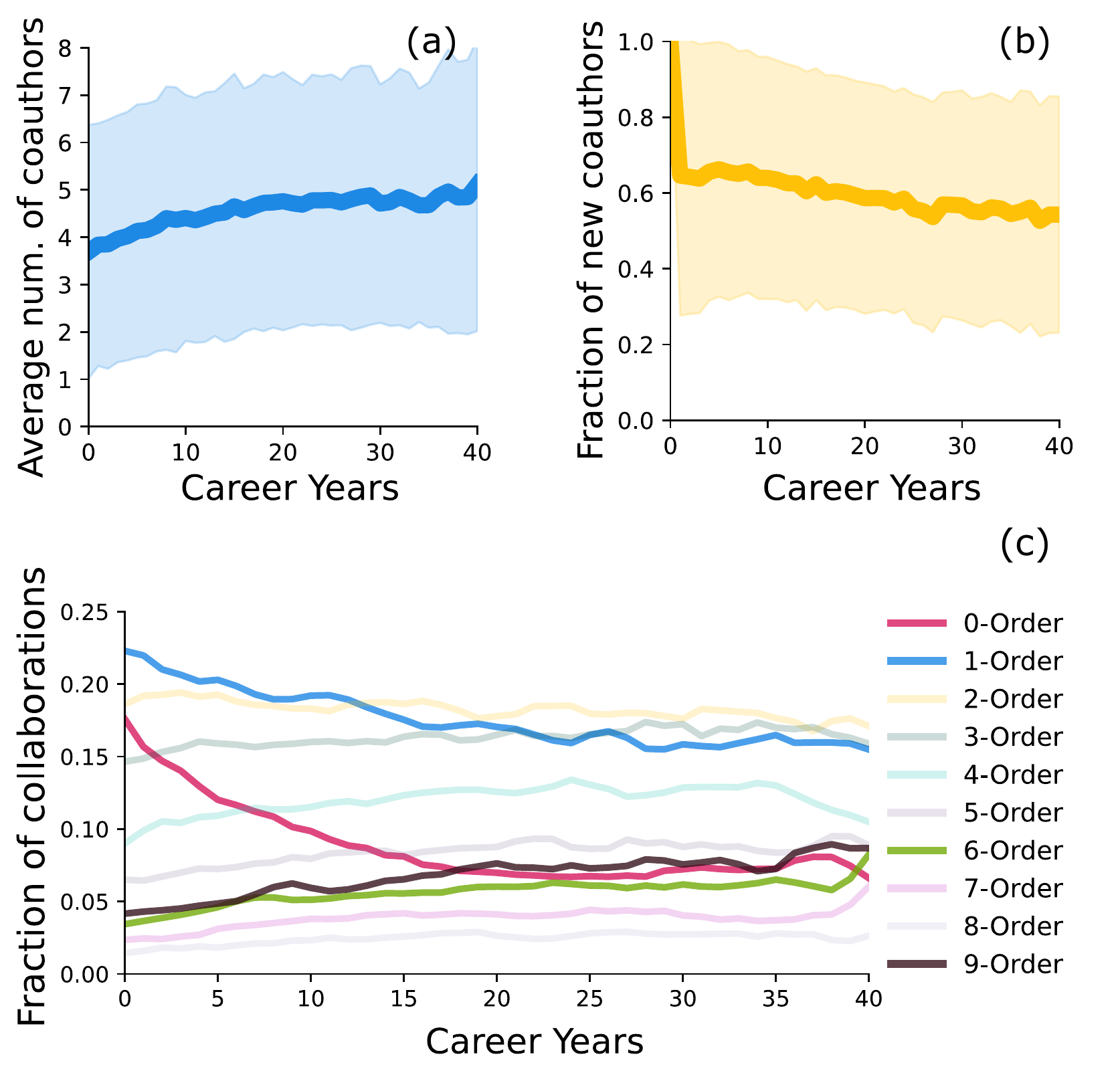}
    \caption{Higher-order analysis of scientific careers. (a) The average number of collaborators per single work over years of career. Scientists work in larger groups the more they progress in their careers. (b) The relative number of new coauthors with respect to the number of coauthors per single year of career. Scientists slightly reduce the chances to collaborate with new coauthors the more they advance in their careers. Solid lines show average values while shaded bands represent one standard deviation confidence intervals. (c) Fraction of collaborations per order over years of career. Scientists are more likely to participate in collaborations of smaller size at the early stage of their careers. The fraction of small collaborations reduces over time. On the contrary, the fraction of large collaborations increases during the career. }
    \label{fig:panel_2}
\end{figure}

From a node-wise analysis, we observe that the average order of collaboration of a single author increases over their career, see Figure~\ref{fig:panel_2}(a). This mesoscale temporal trend is likely to be caused by the different roles played by a scientist during their career. Reasonably, old researchers are likely to be team leaders and part of bigger projects while early-stage scientists often work with their supervisors for their trainees. Moreover, we can't exclude that this pattern can also be influenced by the macroscopic external trend shown in Figure~\ref{fig:panel_1}(a). We hypothesize a coevolution process in which both the technological means and the personal roles evolve during the time span of a career. 
Within the career of an author, we also observe other node-wise patterns in agreement with the previous intuitions. Figure~\ref{fig:panel_2}(c) shows the fraction of works of a single order per career year. The authors manifest a larger propensity to work in small groups at the beginning of their careers while large team projects become more frequent in the last stages of their scientific production. Another possible interpretation is that researchers that do not drop out from academia are the ones that have better skills to work in larger groups.

Another aspect that is worth investigating is how the scientific neighborhood (the set of all coauthors) of a node increases over its career. We, therefore, analyze how prone a researcher is to make new collaborators over the years. Even if a node certainly enlarges its neighborhood over time, the chance that it engages in a higher-order interaction with a node that was never connected to it decreases.
In Figure~\ref{fig:panel_2}(b), we show that, excluding the first year, the fraction of new coauthors within a collaboration decreases linearly over time. This shows a sort of memory attachment mechanism in which nodes preferentially collaborate with nodes with which they have collaborated in the past. Also, this finding is in line with sociological theories concerning the existence of a saturation threshold for an individual's social network neighborhood. Robin Dunbar's theory suggests that a cognitive limit to the number of people with whom one can maintain social relationships exists~\cite{DUNBAR1992469}, and more recently the structure of ego-networks has also been described in terms of Dunbar-cycles which also constraint the number of relations of different kind~\cite{doi:10.1073/pnas.1719233115}. We can imagine that similar thresholds and structures are present for active scientific collaboration connections, which  can be intended as a specific category of social relationships.  

\subsection*{Temporal dynamics of collaborations}
\begin{figure}
    \centering
    \includegraphics[width=\linewidth]{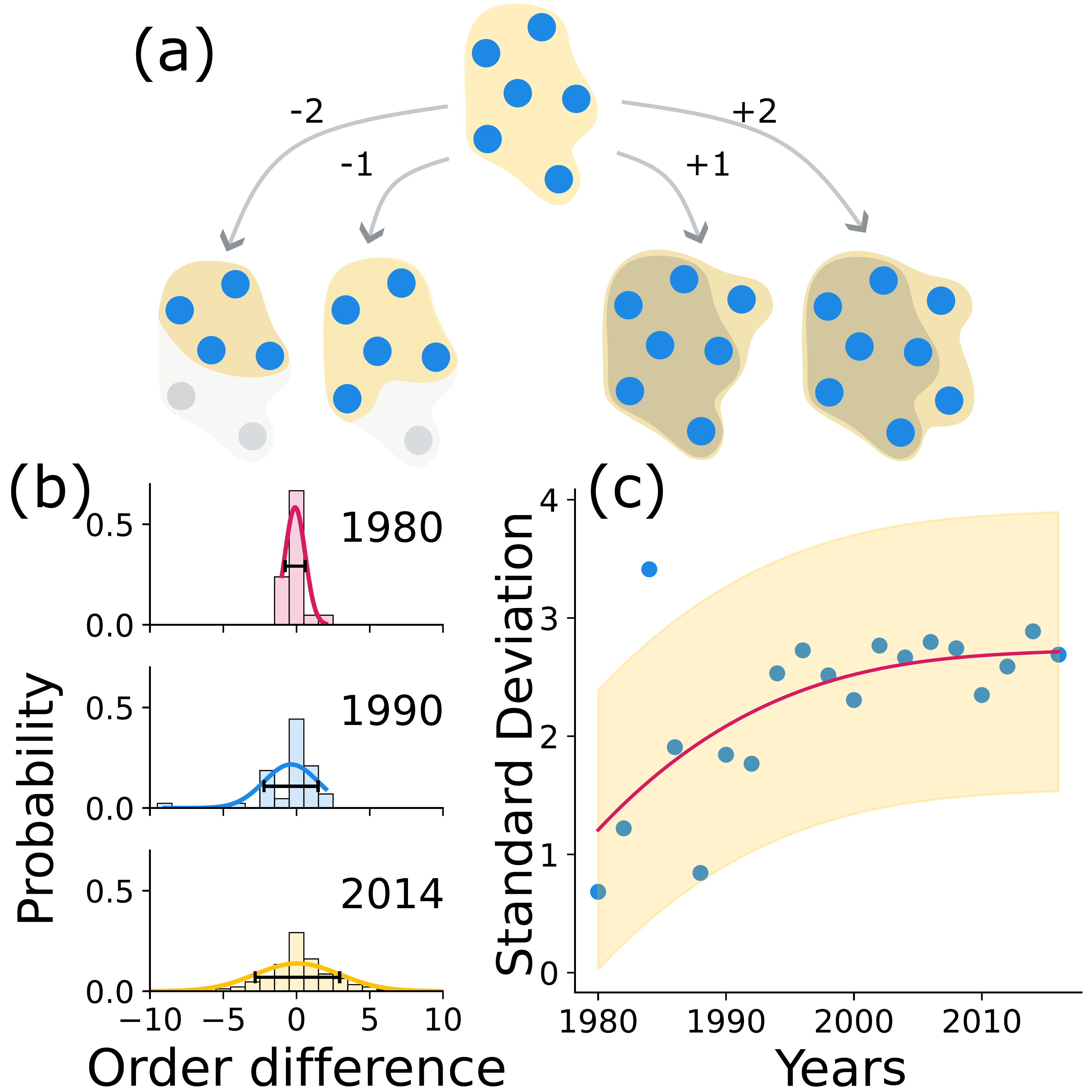}
    \caption{Group dynamics of aggregation and disaggregation. (a) Illustration of aggregation (gaining 1 or 2 nodes) and disaggregation (losing 1 or 2 nodes) processes. (b) Probability of aggregation and disaggregation in scientific collaborations of different decades. Each histogram shows the probability of a group aggregating or disaggregating the next year. $P(x=0)$ is the probability of not changing. $P(x = -n)$ is the probability of disaggregating losing $n$ members. $P(x = n)$ is the probability of aggregating gaining $n$ members. The standard deviation of the probability distribution is a measure of variability in scientific collaboration (solid black line). (c) Standard deviation of aggregation/disaggregation probability for every year, quadratic fit (red line), and its confidence interval (yellow band). Probability distributions have become wider over decades. }
    \label{fig:panel_3}
\end{figure}

The concept that there may be a memory mechanism underlying author connections motivates us to explore the temporal evolution of scientific collaboration groups~\cite{goh2008burstiness,williams2019effects,williams2022shape,gallo2023higher}. Specifically, we investigate how groups of scientists grow or shrink and what are the chances that they keep collaborating together over time. To this end, we analyze the probability that a hyperlink evolves into a subset or into an extension of itself from one year to the next one. To perform the analysis, we made use of ideas introduced by Iacopini et al.~\cite{iacopini2023temporal}, see \textit{Material and methods} for further details.
In Figure~\ref{fig:panel_3}(a), we illustrate four possible scenarios of aggregation (the hyperlink gains nodes) or disaggregation (the hyperlink loses nodes) dynamics.
In Figure~\ref{fig:panel_3}(b), we present the probability distribution of order differences for three representative years. Firstly, we observe that the distribution is centered in zero, suggesting that groups of authors tend to publish another article in the subsequent year with precisely the same set of authors. The distribution is also symmetric, indicating that the probability of gaining new collaborators is comparable to the probability of losing existing members.
Furthermore, the probability distributions increase the variance over the years, suggesting a growing dynamism in losing and acquiring members. These findings are qualitatively valid for all the analyzed years.
In Figure~\ref{fig:panel_3}(c), we plot the standard deviation of the transition probability distribution for every year showing a solid increasing trend in terms of collaboration dynamism.
Interestingly, the variance in order differences manifests a sharp increase up until the mid-90s, while it appears to have stabilized over the subsequent two decades.
These findings indicate that the changes in collaboration order become increasingly significant as time progresses. This observation suggests that collaboration dynamics are evolving, resulting in research teams experiencing substantial changes in collaboration. This could be attributed to the expanding network of authors and the growing connectivity among them. 

\subsection*{Dynamical stability of collaborations}

\begin{figure}
    \centering
    \includegraphics[width=\linewidth]{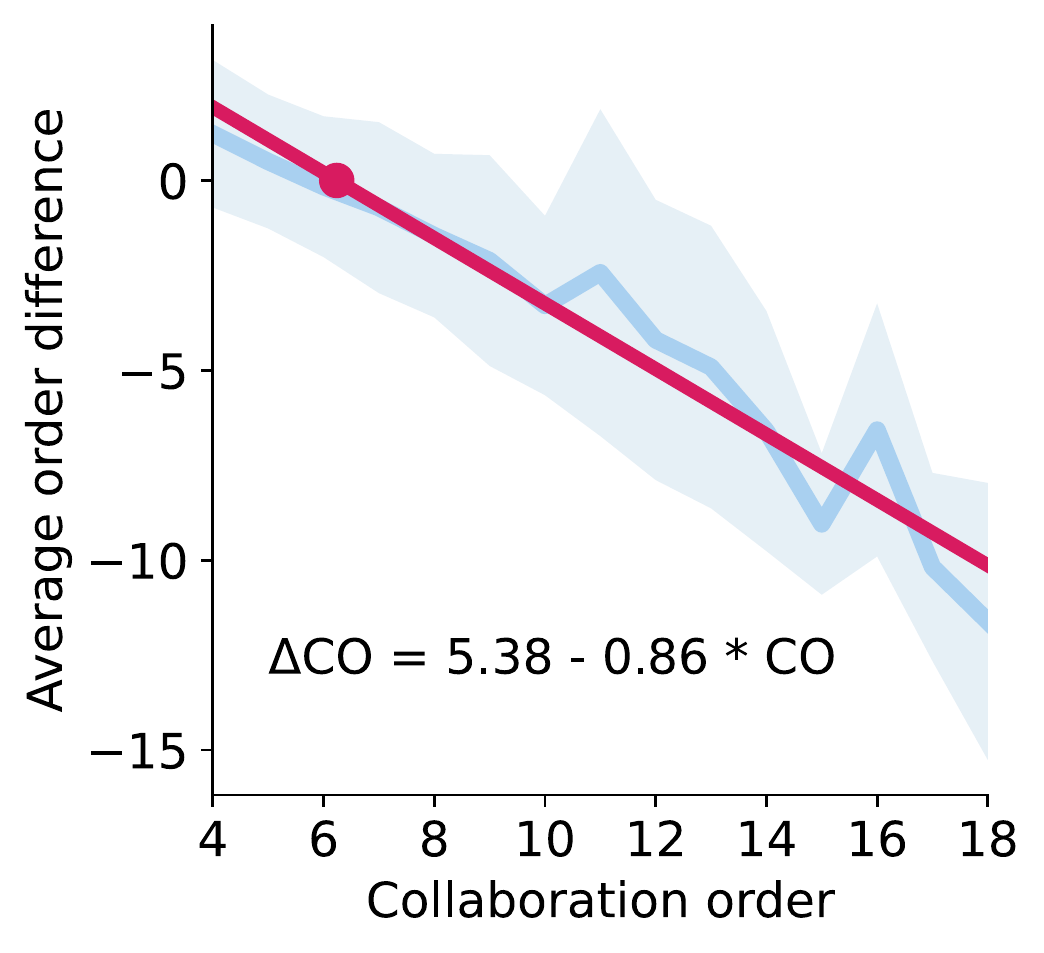}
    \caption{Aggregation trend per order of collaboration. The average variation in collaboration order ($\Delta$CO) during dynamics of aggregation and disaggregation with respect to different orders. The number of members lost (gained) during disaggregation (aggregation) decreases (increases) linearly with respect to the order of collaboration. Small groups increase their members till the fixed point at order 6 (7 members groups), while large groups shrink towards the same point. The parameters of the linear fit are shown in the figures.}
    \label{fig:panel_4}
\end{figure}

Our results suggest increasing dynamism in research collaborations. We further explore how the dynamics of aggregation behave, and how it is influenced by the order of collaborations among authors. In Figure~\ref{fig:panel_4}, we present the average order difference in relation to the collaboration order. Interestingly, we have observed a linear correlation between the order of collaborations and the average order difference. Let $CO$ be the collaboration's order, we call $\Delta CO$ the collaboration's order variation in the following year. Then the empirical law $\Delta CO = 5.38 - 0.86 CO$ holds.

This allows us to identify a specific collaboration order (i.e. $CO^* = 6.24$) that can be interpreted as a fixed point of the group aggregation and disaggregation dynamics. At this order of interaction, there is a balance between groups tending to shrink and those tending to enlarge. That is to say, for orders equal to 6, the average order difference is approximately zero, indicating that groups consisting of 7 authors are the least likely to change their size.
This can be also interpreted as the order in which we have dynamical stability in terms of aggregation mechanism. Indeed, when the collaboration order is above $CO^*$, the average order difference tends to bring the order to that value. On the contrary, when the collaboration order is below $CO^*$, the average order difference is positive. Also, being the absolute value of the \textit{rate} below 1 ($\sim 0.86$), collaborations tend to $CO^*$ but do not reach it in a single time step (one year). At the following time step, it is likely that the group will continue shrinking (or enlarging), eventually reaching $CO^*$. This gradual convergence implies that $CO^*$ behaves as an asymptotically stable fixed point.

As a robustness check, we perform a stratified analysis by dividing the data into different time windows. Remarkably, our results demonstrate a consistent pattern across these distinct time intervals. Specifically, we observe a consistent linear correlation between the average order difference and the collaboration order, with comparable angular coefficients. This suggests that the observed relationship is robust and persists over time, regardless of the specific time window under consideration.

%maybe this part goes to discussion
By verifying the consistency of the linear correlation across various time windows, we strengthen the validity and generalizability of our findings. 
This provides further evidence to support the notion that the collaboration order has a significant impact on the average order difference and, consequently, on the dynamics of group size changes.
The identified fixed point and the observed linear correlation provide insights into the stability and behavior of research groups in terms of their aggregation mechanism.

\section*{Discussion}
In this study, we have analyzed human collaboration on a co-authorship dataset by means of a higher-order perspective. Our focus is on identifying temporal higher-order patterns at the global, group, and node levels, as well as exploring the dynamic characteristics of real group dynamics. Our comprehensive analysis provides valuable insights into a more detailed representation of the temporal properties of collaborations among human beings. Furthermore, our study offers novel data analysis methods for evaluating temporal hypergraph dynamics, an area lacking research thus far.

% replicate to improve applicability of results
To confirm the applicability of our proposed approach and establish the generality of the observed patterns, it is crucial to replicate the analysis on additional co-authorship datasets. Furthermore, extending the analysis to other forms of human interactions and collaborations would provide a broader perspective. 

% other applications
Our findings suggest that the proposed method can be effectively applied in various scenarios where monitoring the temporal evolution of groups is desired, enabling quantification of the tendency of these groups to aggregate or disaggregate, ultimately reaching a specific equilibrium size. For instance, in ecology~\cite{montoya2006ecological,fath2007ecological}, our approach can aid in quantifying population dynamics over time. In transportation~\cite{estrada2008communicability}, it can assist in measuring changes in the mobility patterns of groups of individuals. Moreover, this methodology can be employed in engineering to track groups of devices engaged in communication and collaboration within distributed applications, such as fleets of unmanned vehicles or wireless sensor networks~\cite{lv2002search, cao2012overview, elhoseny2019dynamic}.
Importantly, this approach holds broader applicability in studying the evolution of phenomena where groups of elements undergo temporal changes. For example, in the realm of language evolution~\cite{sowa1992semantic,lehmann1992semantic}, our method can provide insights into how groups of words may initially exhibit proximity but gradually disperse into different clusters as the language evolves, which can add a temporal dimension to natural language processing analysis.

%limitation
It is important to note that while our method effectively detects groups that gain or lose a specific number of members, it does not capture partial overlaps, which are common in co-authorship networks as well as the other applications we mentioned. To address this limitation, we suggest the computation of metrics capable of quantifying partial overlap and conducting the analysis by considering groups that exceed a certain overlap threshold, providing a more nuanced understanding of group dynamics. Furthermore, our study compares each year with the subsequent year, assuming that groups typically publish at least one paper annually. However, it would be valuable to replicate the analysis using finer or coarser temporal resolutions to explore whether different time ranges capture the same or other phenomena. Fine-tuning the temporal analysis can offer insights into the stability or dynamism of groups at various time scales. Therefore, when applying this approach, it is crucial to carefully consider the appropriate time scale that aligns with the specific phenomenon under investigation. Lastly, while our proposed method offers valuable insights into the temporal dynamics of group collaborations, it does not provide causal explanations for the observed patterns. Future research could delve into the underlying mechanisms driving these dynamics and explore the factors influencing group formation, dissolution, and evolution over time.

In conclusion, our study makes a valuable contribution to advancing our understanding of human collaboration dynamics, encompassing global, group, and personal characteristics. The potential of our proposed method reaches far beyond the realm of co-authorship networks. It offers prospects for investigating diverse domains and providing valuable insights into the evolution of groups across various contexts.

\section*{MATERIALS AND METHODS}

\subsection*{Data Description}
The dataset used in this study is a temporal higher-order network dataset. It consists of a sequence of timestamped simplices, where each simplex represents a publication from the "Geology" category in the Microsoft Academic Graph~\cite{landry2023xgi}. In this context, nodes within the simplices represent authors, and the timestamps correspond to the year of publication. The dataset includes 1,256,385 nodes representing authors and 1,590,335 timestamped simplices representing publications.

%\subsection*{Dynamical stability}

\subsection*{Temporal dynamics of collaborations}

We conduct an analysis of the temporal dynamics of simplices. Specifically, we examine how groups of authors involved in an article in a given year either add or lose members in the following year. To perform this analysis, we focus on medium-sized simplices and limit the dataset to simplices ranging in size from 5 to 20 authors. Within this dataset, we examine the simplices that either include or are included by other groups in the subsequent year and we calculate the order difference between these groups.
To detect the processes of disaggregation, we examine simplices in a given year and identify the presence of lower-order simplices in the following year. Similarly, to detect aggregating processes, we examine simplices in the subsequent year and determine the presence of lower-order simplices in the current year. We also record the disparity in order between these simplices.
For instance, if a group of authors published an article in year Y, and the same group published another article in year Y+1 with the addition of $n$ members, we record this instance with an order difference of $n$. If the exact same group of researchers published two papers in consecutive years, we record an order difference of 0. On the other hand, if a subset that did not include $n$ members from the original group published a paper in the following year, we consider an order difference of $-n$. Therefore, negative and positive order differences indicate the disaggregation and aggregation of groups, respectively. This process is illustrated in Figure~\ref{fig:panel_3}a.

To investigate the temporal progression of this phenomenon, we compute the order differences for alternate years spanning from 1980 to 2016. Preceding 1980, the group sizes were small on average, and few simplices of the considered order and displaying (dis)aggregating properties could be identified. In two separate analyses, we stratify the computed order differences based on the year and order size. In both analyses, we evaluate the probability distribution of order differences for each sample, as well as its mean and standard deviation. 

In the temporal analysis, we fit the non-linear relationship between time and the standard deviation of the order difference with a polynomial function of order 3. In the size-based analysis, we conduct a regression analysis to identify the best-fit line that captures the relationship between the average order difference and the collaboration order. To check the robustness of this linear trend, we perform additional investigations by stratifying the data based on different time windows. Specifically, we divide the dataset into consecutive windows of 10 years, spanning from 1980 to 2016. Within each window, we examine the average order difference and its relationship with the collaboration order. The linear fit is consistent and robust across time window size.

%This information can help understand the trends and patterns in research collaboration, providing valuable knowledge about the stability, growth, and fragmentation of research groups.

%This approach allows one to track the changes in group composition over time and understand how authors collaborate and form new bigger groups or disperse among existing groups. By examining the relationships between groups from one year to the next, we can gain insights into the dynamics of research collaborations and the patterns of authorship within the analyzed dataset.

%\section{\label{sec:model} ACTIVITY DRIVEN MODEL}

%To describe the observed analysis from the data, we use an activity driven model.....In this model we assume an intial reservoir of authors that will create collaborations at a certain rate proportional to the number of collaborations from the previous year. Collaborations will be created by a "leader node", which is selected randomly according to an activation probability $p_A(t)$. This activation probability is monotonically with %t%, inspired from the data. 

\subsection*{Data and Code Availability}
The dataset used in this study is available for download at \url{https://www.cs.cornell.edu/~arb/data/coauth-MAG-Geology/}. The code to reproduce the analysis and generate the figures is provided in the GitHub repository \url{https://github.com/juanfernandezgracia/complexity_72h}. Part of the analysis was conducted using the package HGX~\cite{lotito2023hypergraphx}.

%\columnbreak
\section*{ACKNOWLEDGEMENTS }

This work is the output of the Complexity72h workshop, held at IFISC in Palma, Spain, 26-30 June 2023 (\verb!https://www.complexity72h.com!). We acknowledge I. Iacopini and A. Galeazzi, for the fruitful discussion and suggestions for this work. We acknowledge the Organizing Commitee and all participants of the workshop for contributing to an inspinring and interactive work environment.
%\cite{cencetti2021temporal,kim2022higher,benson2018simplicial,xie2021distributed,fortunato2018science,ubaldi2017burstiness,vasilyeva2021multilayer,xie2016geometric,newman2001structure,wang2021science,newman2018networks,lung2018hypergraph,barabasi2002evolution,Patania2017TheSO}. 

\bibliography{Bibliography}% Produces the bibliography via BibTeX.

\end{document}